\documentclass[%
 reprint,
 amsmath,amssymb,
 aps,
]{revtex4-2}

\usepackage{graphicx}
\usepackage{dcolumn}
\usepackage{bm}
\usepackage{braket}
\usepackage{hyperref}

\begin{document}

\preprint{APS/123-QED}

\title{Telecom-compatible cross-band quantum memory via dual photon modes dark-state polaritons}

\author{Dounan Du}
 \email{dounan.du@stonybrook.edu}
\affiliation{%
Department of Physics and Astronomy, Stony Brook University, Stony Brook, 11794-3800, NY, USA\\
}%
\author{Eden Figueroa}%
 \email{eden.figueroa@stonybrook.edu}
\affiliation{%
Department of Physics and Astronomy, Stony Brook University, Stony Brook, 11794-3800, NY, USA\\
}%
\affiliation{%
Brookhaven National Laboratory, Upton, 11973-5000, NY, USA\\
}%

\date{\today}

\begin{abstract}
Quantum memories are essential components of quantum networks, enabling synchronization, quantum repeaters, and long-distance entanglement distribution. Most ensemble-based realizations rely on dark-state polaritons (DSPs) in $\Lambda$-type systems that operate at near-infrared wavelengths, such as 795 nm in $^{87}$Rb, far from the telecom band where long fiber transmission is optimal. Here we identify a DSP in $^{87}$Rb that coherently couples two photonic modes at 795 nm and 1324 nm through a shared spin-wave coherence. We derive its field operator and group velocity, extending the Fleischhauer–Lukin model to a dual-wavelength regime, and formulate a memory protocol enabling bidirectional storage and retrieval between the two modes. Numerical simulations of the full six-level dynamics confirm two-way storage and retrieval for both same-mode and cross-mode operation between the two wavelengths. The results demonstrate a dual-wavelength memory that unifies node-band and telecom-band operation within a single ensemble, providing a potential route toward frequency-conversion-free quantum-network interfaces.
\end{abstract}

\maketitle

\textit{Introduction.—}Quantum memories are essential components of emerging quantum network technologies \cite{kimble2008quantum, heshami2016quantum, azuma2023quantum, lu2022micius, liu2024creation}. They enable long-distance entanglement distribution \cite{duan2001long, knaut2024entanglement, zhou2024long, liu2024creation}, synchronization of probabilistic processes\cite{heshami2016quantum, davidson2023single}, and the implementation of quantum repeater protocols \cite{sangouard2011quantum, azuma2023quantum}. In fiber-based architectures, memories provide the crucial interface between stationary matter systems and traveling photonic qubits, making them central to scalable quantum communication \cite{kimble2008quantum, lvovsky2009optical, bao2012quantum, liu2024creation}.

Dark-state polaritons have been studied extensively in quantum optics since their first identification \cite{fleischhauer2000dark, phillips2001storage}, and have become the foundation of ensemble-based quantum memories \cite{lvovsky2009optical, ma2017optical}. In the canonical $\Lambda$-type three-level configuration, an optical probe can be coherently mapped onto a collective spin excitation between two metastable ground states by adiabatically reducing the control field, enabling reversible storage and retrieval of single photons with high fidelity \cite{fleischhauer2000dark, lvovsky2009optical,heshami2016quantum}. This mechanism has been demonstrated across a wide variety of platforms and remains the cornerstone of quantum memory implementations for quantum networking \cite{chen2008memory, chaneliere2005storage, appel2008quantum}.

Despite these advances, most atomic-ensemble memories operate at near-infrared wavelengths determined by internal transitions, such as 795 nm in $^{87}$Rb. These wavelengths are far from the telecom bands where fiber transmission is optimal, posing a major challenge for long-distance quantum networks \cite{gisin2007quantum}. Bridging this gap typically requires external frequency conversion \cite{radnaev2010quantum, wengerowsky2025quantum}, but conversion adds loss, noise, and technical complexity \cite{radnaev2010quantum}. Telecom-band memories have also been explored \cite{liu2022demand, thomas2024deterministic, thomas2023single}, yet they still require conversion to interface with near-infrared quantum nodes. Thus, the absence of a memory that natively supports both node-compatible and telecom photons remains a practical hurdle, forcing stand-alone memories to rely on frequency conversion for deployment in large-scale quantum networks.

In this work, we propose a multi-level atomic configuration that supports a dark-state polariton coherently linking two photonic modes of widely separated wavelengths. We identify the conditions under which this dual-mode polariton emerges and formulate a quantum memory protocol that exploits it for reversible mapping between node-band and telecom-band photons. As a concrete realization, we simulate the full six-level dynamics of $^{87}$Rb, where the relevant optical transitions occur at 795 nm and 1324 nm. The numerical results confirm all storage–retrieval pathways—same-mode operation at each wavelength and cross-mode conversion between the two, thereby demonstrating the feasibility of cross-band quantum storage and retrieval within a single atomic ensemble.

\begin{figure}
\includegraphics{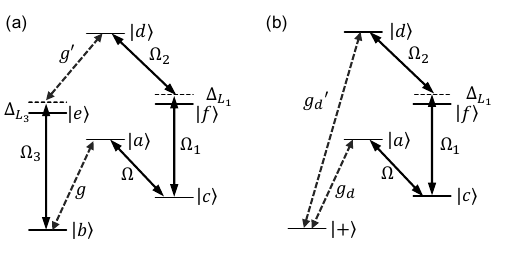}
\caption{\label{fig:energy_level} (a) Atomic level structure supporting a dual-photon-wavelength dark-state polariton. Two $\Lambda$-type subsystems are coherently connected by control fields: $\Omega_{3}$ couples $\ket{e}$ and $\ket{b}$, and $\Omega_{1}$ couples $\ket{f}$ and $\ket{c}$. The lower $\Lambda$ is coupled to a classical control field $\Omega$ and a quantum probe field with coupling constant $g$, while the upper $\Lambda$ is coupled to a control field $\Omega_{2}$ and a quantum field with coupling constant $g'$. (b) Effective level structure of (a). Both quantum fields $g$ and $g'$ are coupled to the dressed state $\ket{+}$ formed from $\ket{e}$, $\ket{b}$ and field $\Omega_3$.}
\end{figure}

\textit{Dark-state polaritons with dual photon modes arises from dressed state.—}
We consider an atomic medium comprising two connected $\Lambda$-type level-structure with two metastable ground states, as shown in Fig.~\ref{fig:energy_level}(a). A quantum prob field of mode $\hat{a}$ couples the metastable state $\ket{b}$ and excited state $\ket{a}$ with coupling constant $g$, forming the lower $\Lambda$-type level-structure. A second quantum field of mode $\hat{b}$ couples excited states $\ket{e}$ and $\ket{d}$ with coupling constant $g'$, forming the prob field of the higher $\Lambda$-type level-structure. The two $\Lambda$-type level-structures are connected by strong control fields $\Omega_1$ and $\Omega_3$, which couple the ground states $\ket{b}$ and $\ket{c}$ to the excited states $\ket{e}$ and $\ket{f}$, respectively. The energy structure resembles double-$\Lambda$ type media with shared metastable ground states, in which dark state polaritons with two photonic components have been identified \cite{chong2008dark}. 

The coupling of levels $\ket{e}$ and $\ket{b}$ by the coherent driving field $\Omega_3$ connects the two $\Lambda$ systems, forming a dressed-state manifold that serves as an effective shared ground state for both quantum fields. In rotating frame the sub-system of level $\ket{e}$, $\ket{b}$ and field $\Omega_3$ is governed by Hamiltonian
\begin{equation}
\begin{aligned}
\tilde{H}_{sub} &=\hbar\sum_r[-\Delta_{L_3}\sigma_r^{ee}] -(\hbar \Omega_3 \sum_re^{ik_{L_3}r}\sigma_r^{eb}+\text{H.c.})\\
\end{aligned}
\end{equation}
where $\sigma_{r}^{ij}=\ket{i}_r \bra{j}_r$ is the atomic operator at position $r$ and $\Delta_{L_3}=\omega_{L_3}-\omega_{eb}$, which we assume a far de-tuned $\Delta_{L_3}>0$. We assume the atoms are initially in the ground state $\ket{b}$, and $\Omega_3$ is turned on adiabatically. The atoms thus result in eigen-state
\begin{equation}
\begin{aligned}
\ket{+}_r&=\sin\theta \ket{b}_r+e^{ik_{L_3}r}\cos \theta \ket{e}_r 
\end{aligned}  
\end{equation}
with energy 
\begin{equation}
\begin{aligned}
E_{+} &=\hbar(-\Delta_{L_3}+\sqrt{\Delta^2_{L_3}+4\Omega_3^2})/2,
\end{aligned}  
\end{equation}
where $\tan 2\theta = 2\Omega_3/\Delta_{L_3}$. The system thus reduces to the effective level structure shown in Fig.~\ref{fig:energy_level}(b), in which both quantum fields $\hat{a}$ and $\hat{b}$ couple to the dressed state $\ket{+}$. 

The rotating-frame effective Hamiltonian of the full system can be written as the sum of free and interaction terms. The interaction-free Hamiltonian is
\begin{equation}
\begin{aligned}
\tilde{H}_0 &=\hbar\sum_r[(-\Delta_L+\omega_c-\frac{E_+}{\hbar})\sigma_r^{aa}+(\omega_c-\frac{E_+}{\hbar})\sigma_r^{cc}\\
&+(-\Delta_{L_1}-\Delta_{L_2}+\omega_c-\frac{E_+}{\hbar})\sigma_r^{dd}\\
&+(-\Delta_{L_1}+\omega_c-\frac{E_+}{\hbar})\sigma_r^{ff}]\\
&+\sum_{k_1}\hbar (c|k_1|-\omega_L)a^{\dagger}_{k_1}a_{k_1}\\
&+\sum_{k_2}\hbar [c|k_2|-(\omega_{L_1}+\omega_{L_2}-\omega_{L_3})]b^{\dagger}_{k_2}b_{k_2}\\
\end{aligned}
\end{equation}
while the interaction between the atoms and the two quantum fields is governed by
\begin{equation}
\begin{aligned}
\tilde{H}_1 = &-\hbar g_d\sum_{k_1}a_{k_1} \sigma_{k_1}^{a+}-\hbar g_d'\sum_{k_2}b_{k_2} \sigma_{k_2+k_{L_3}}^{d+}+\text{H.c.}\\
\end{aligned}
\end{equation}
where the collective atomic spin operators are defined as $\sigma_{k}^{ij}=\frac{1}{\sqrt{N}}\sum_r\ket{i}_r \bra{j}_r e^{ikr}$, $g_d=g\sin\theta$ and $g_d'=g'\cos\theta$. N is the number of atoms inside the field mode volume. Additional interaction terms arise from the coupling to the control fields
\begin{equation}
\begin{aligned}
\tilde{H}_2 =
&-\hbar \Omega \sum_re^{ik_Lr}\sigma_r^{ac} -\hbar \Omega_1 \sum_re^{ik_{L_1}r}\sigma_r^{fc}\\
&-\hbar \Omega_2 \sum_re^{ik_{L_2}r}\sigma_r^{df}+\text{H.c.}\\
\end{aligned}
\end{equation}

We seek quantum field mode operators $\Psi_k^{\dagger}$ that satisfy commutation relations
\begin{equation}
\begin{aligned}
\ [\tilde{H}, \Psi_k^{\dagger}] = \hbar\omega_k \Psi_k^{\dagger}+\ ...\\
\label{eq:eigen}
\end{aligned}
\end{equation}
in weak excitation limit, and represent a linear superposition of atomic excitation operators and photon creation operators
\begin{equation}
\begin{aligned}
\Psi_k^{\dagger} &= \phi_k^1\sigma^{a+}_{k+k_L} + \phi_k^2\sigma^{c+}_{k} +\phi_k^3\sigma^{f+}_{k+k_{L_1}}+\phi^4_k\sigma^{d+}_{k+k_{L_1}+k_{L_2}}\\
&+\phi^5_k a_{k+k_L}^{\dagger}
+\phi^6_k b_{k+k_{L_1}+k_{L_2}-k_{L_3}}^{\dagger}.
\end{aligned}
\end{equation}
The neglected terms in the commutation relations vanish when acting on the ground state. Acting $\Psi_k^{\dagger}$ on the ground state thus generates an eigenstate of the coupled system. We use the following commutations of the atomic excitation operators and photon creation operators
\begin{widetext}
\begin{equation}
\begin{aligned}
\ [\tilde{H}, \sigma_{k+k_L}^{a+}] &= \hbar(-\Delta_L+\omega_c-\frac{E_+}{\hbar})\sigma_{k+k_L}^{a+}-\hbar\Omega^*\sigma^{c+}_{k}-\hbar g_d^*\sum_{k_1}a^{\dagger}_{k_1}I_{k+k_L-k_1}\ +\ ...\\
[\tilde{H}, \sigma_{k}^{c+}] &= \hbar(\omega_c-\frac{E_+}{\hbar})\sigma^{c+}_k-\hbar\Omega\sigma^{a+}_{k+k_L}-\hbar\Omega_1\sigma^{f+}_{k+k_{L_1}}+\ ...\\
[\tilde{H}, \sigma_{k+k_{L_1}}^{f+}] &= \hbar(-\Delta_{L_1}+\omega_c-\frac{E_+}{\hbar})\sigma_{k+k_{L_1}}^{f+}-\hbar\Omega^*_1\sigma_{k}^{c+}-\hbar\Omega_2\sigma_{k+k_{L_1}+k_{L_2}}^{d+}+\ ...\\
[\tilde{H}, \sigma_{k+k_{L_1}+k_{L_2}}^{d+}] &=\hbar(-\Delta_{L_1}-\Delta_{L_2}+\omega_c-\frac{E_+}{\hbar})\sigma^{d+}_{k+k_{L_1}+k_{L_2}}-\hbar\Omega^*_2\sigma_{k+k_{L_1}}^{f+}-\hbar g_d'^*\sum_{k_2}b^{\dagger}_{k_2}I_{k+k_{L_1}+k_{L_2}-k_2-k_{L_3}}\ +...\\
[\tilde{H}, a^{\dagger}_{k+k_L}] &= \hbar c|k|a^{\dagger}_{k+k_L}-\hbar g_d\sigma^{a+}_{k+k_L}\\
[\tilde{H}, b^{\dagger}_{k+k_{L_1}+k_{L_2}-k_{L_3}}] &=\hbar c|k|b^{\dagger}_{k+k_{L_1}+k_{L_2}-k_{L_3}}-\hbar g_d'\sigma^{d+}_{k+k_{L_1}+k_{L_2}}.\\
\end{aligned}
\end{equation}
\end{widetext}
Under the collinear phase match condition $k+k_L-k_1=0$ and $k+k_{L_1}+k_{L_2}-k_2-k_{L_3}=0$, and further assuming $\Delta_L=0$ and $\Delta_{L_1}+\Delta_{L_2}=0$, Eq.~(\ref{eq:eigen}) reduces to the eigen-equation
\begin{equation}
\begin{aligned}
&
\begin{pmatrix}
\omega_c' & -\Omega^* & 0 & 0 & -g_d^* & 0\\
-\Omega & \omega_c' & -\Omega_1 & 0 & 0 & 0\\
0 & -\Omega_1^* & -\Delta_{L_1}+\omega_c' & -\Omega_2 & 0 & 0 \\
0 & 0 & -\Omega_2^* & \omega_c' & 0 & -g_d'^* \\
-g_d & 0 & 0 & 0 & c|k| & 0 \\
0 & 0 & 0 & -g_d' & 0 & c|k| \\
\end{pmatrix}
\begin{pmatrix}
\phi_k^1\\
\phi_k^2\\
\phi_k^3\\
\phi_k^4\\
\phi_k^5\\
\phi_k^6\\
\end{pmatrix}\\
&=\omega
\begin{pmatrix}
\phi_k^1\\
\phi_k^2\\
\phi_k^3\\
\phi_k^4\\
\phi_k^5\\
\phi_k^6\\
\end{pmatrix}\\
\end{aligned}
\end{equation}
where $\omega_c'=\omega_c-\frac{E_+}{\hbar}$. We focus on eigenmodes with zero or negligible excitation to levels susceptible to spontaneous emissions ($\phi^1$, $\phi^3$, $\phi^4$) and dominant excitation to metastable state or photon modes ($\phi^2$, $\phi^5$, $\phi^6$). For $\Delta_{L_1}>0$, a solution
\begin{equation}
\begin{aligned}
    \Psi_k^{\dagger} &\propto \sigma_k^{c+}-\frac{\Omega_1}{\Delta_{L_1}}\sigma_{k+k_{L_1}}^{f+}-\frac{\Omega^*}{g_d^*} a_{k+k_L}^{\dagger} \\
    &+ \frac{\Omega_2^*\Omega_1}{\Delta_{L_1}g_d'^*}b_{k+k_{L_1}+k_{L_2}-k_{L_3}}^{\dagger}
\end{aligned}
\end{equation}
to Eq.~(\ref{eq:eigen}) exists when pump field $L_1$ is far detuned $|\Omega_1| \ll \Delta_{L_1}$. 

The eigenmode represents a hybrid light–matter excitation that simultaneously couples two distinct photonic modes to a shared atomic spin wave, forming the basis for dual-wavelength quantum memory operation. In weak excitation limit, the field mode operators satisfy the bosonic commutation relations
\begin{equation}
\begin{aligned}
    \ [\Psi_{k'}, \Psi_k^{\dagger}] = \delta_{k',k}A^2[1&+(\frac{\Omega_1}{\Delta_{L_1}})^2+(\frac{\Omega^*}{g_d^*})^2\\
    &+(\frac{\Omega_2^*\Omega_1}{\Delta_{L_1}g_d'^*})^2]+...
\end{aligned}
\end{equation}
where $A$ is a normalization factor ensuring unity of commutation for $k=k'$. Terms neglected in the commutation relation vanish when acting on the vacuum. The group velocity of the polariton field is 
\begin{equation}
\begin{aligned}
    v_g 
    &= c\cdot\frac{(\frac{\Omega^*}{g_d^*})^2+(\frac{\Omega_2^*\Omega_1}{\Delta_{L_1}g_d'^*})^2}{1+(\frac{\Omega_1}{\Delta_{L_1}})^2+(\frac{\Omega^*}{g_d^*})^2+(\frac{\Omega_2^*\Omega_1}{\Delta_{L_1}g_d'^*})^2}.
\label{eq:groupv}
\end{aligned}
\end{equation}
Acting on the vacuum, the mode operator creates a dark-state polariton comprising two photonic components corresponding to modes $\hat{a}$ and $\hat{b}$, and an atomic spin-wave component. The photonic contributions are proportional to the Rabi frequencies of the control fields, $\Omega$ for mode $\hat{a}$ and $\Omega_1\Omega_2$ for mode $\hat{b}$. By adiabatically varying the control fields, the relative amplitudes of the photonic and spin-wave components can be continuously tuned, enabling conversion between photon and atomic excitations. In particular, Eq.~(\ref{eq:groupv}) shows that the group velocity vanishes when $\Omega=\Omega_1\Omega_2=0$, corresponding to complete storage of the optical excitation in the atomic ensemble.

\begin{figure}[t]
\includegraphics{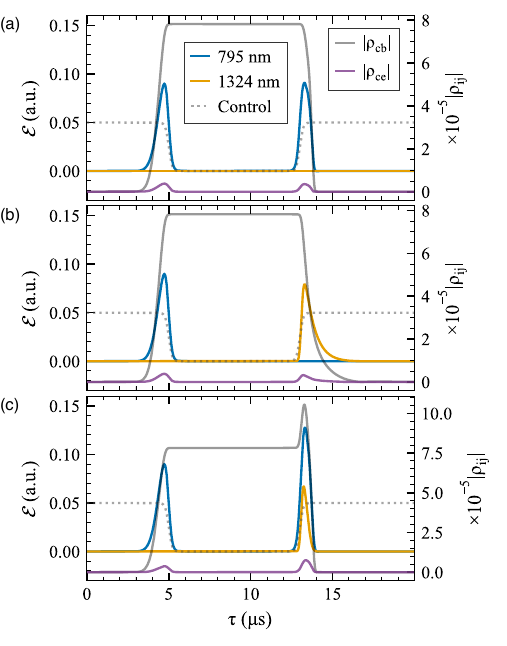}
\caption{\label{fig:a_in} Numerical simulation of storage and retrieval of a 795 nm probe field in a $^{87}$Rb ensemble of length 1.4 cm. Field amplitudes and selected density-matrix elements at the output end ($z=1.4$ cm) are plotted as a function of the retarded time $\tau=t-z/c$. (a) A Gaussian probe pulse centered at $\tau_{0}=4.3~\mu\mathrm{s}$ enters the ensemble while continuous-wave control fields $\Omega_{3}$ and $\Omega$ follow the dashed temporal profile and are turned off at $\tau_{\mathrm{off}}=5~\mu\mathrm{s}$, then reactivated at $\tau_{\mathrm{on}}=13~\mu\mathrm{s}$. (b) Same as (a), but at $\tau_{\mathrm{on}}$ the control fields $\Omega_{3}$, $\Omega_{1}$, and $\Omega_{2}$ are applied according to the same profile. (c) Same as (a), with $\Omega_{3}$, $\Omega$, $\Omega_{1}$, and $\Omega_{2}$ all turned on during retrieval. Field amplitudes are in arbitrary units; dashed curves indicate the control-field time profiles, whose scales are not to be compared with those of the probe fields. The results illustrate controlled storage and cross-band retrieval of the 795 nm photon through different combinations of control fields.}
\end{figure}

\textit{Dual-band quantum memory.—} To verify the dual-band quantum memory protocol predicted above, we perform numerical simulations based on the full six-level structure of $^{87}$Rb, shown in Fig.~\ref{fig:energy_level}(a). The relevant states are $\ket{b}=\ket{5S_{1/2},F=1}$, $\ket{c}=\ket{5S_{1/2},F=2}$, $\ket{a}=\ket{5P_{1/2},F=1}$, $\ket{e}=\ket{5P_{1/2},F=2}$, $\ket{f}=\ket{5P_{3/2},F=1}$, and $\ket{d}=\ket{6S_{1/2},F=1}$. The probe fields $\hat{a}$ and $\hat{b}$ are treated as weak classical envelopes $\mathcal{E}(z,t)$ governed by the one-dimensional propagation equation,
\begin{equation}
\begin{aligned}
(\frac{1}{c}\frac{\partial}{\partial t}+\frac{\partial}{\partial z})\mathcal{E}(z, t) &= \frac{ik}{\epsilon_0}nd_{ml}\tilde{\rho}_{lm}(z,t) 
\end{aligned}
\end{equation}
where $\tilde{\rho}_{lm}(z,t)$ denotes the density matrix element between states $\ket{l}$ and $\ket{m}$, $d_{ml}$ the dipole moment and $n$ the atomic density. The atomic evolution is governed by the master equation
\begin{equation}
\begin{aligned}
\frac{d\tilde{\rho}}{dt} &= -\frac{i}{\hbar} [\tilde{H}, \tilde{\rho}] + \sum_j \left[ 2L_j \tilde{\rho} L_j^\dagger - \{L_j^\dagger L_j, \tilde{\rho}\} \right]
\end{aligned}
\end{equation}
where $\tilde{H}$ includes all optical couplings among the six levels. Numerical simulations are performed for experimentally accessible parameters corresponding to a $^{87}$Rb ensemble of length 1.4 cm (see Supplemental Material for details on parameters and numerical integration~\cite{SM}).

\begin{figure}[t]
\includegraphics{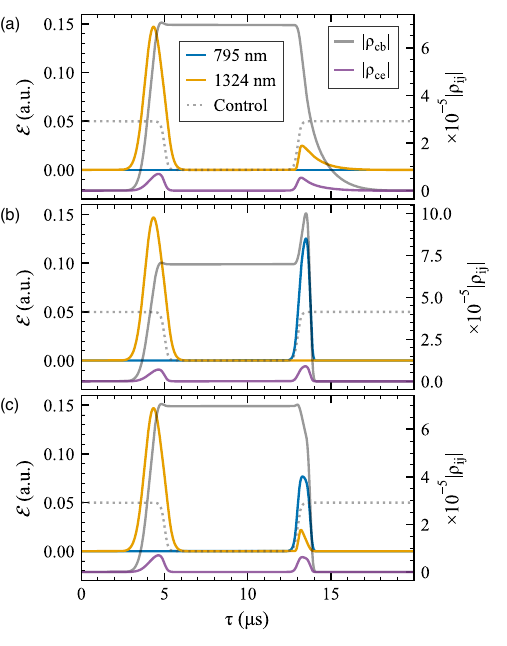}
\caption{\label{fig:b_in} Numerical simulation of storage and retrieval of a 1324 nm probe field in a $^{87}$Rb ensemble of length 1.4 cm. Field amplitudes and selected density-matrix elements at the output end ($z=1.4$ cm) are plotted as a function of the retarded time $\tau=t-z/c$. (a) A Gaussian probe pulse centered at $\tau_{0}=4.3~\mu\mathrm{s}$ enters the ensemble while continuous-wave control fields $\Omega_{3}$, $\Omega_1$ and $\Omega_2$ follow the dashed temporal profile and are turned off at $\tau_{\mathrm{off}}=5~\mu\mathrm{s}$, then reactivated at $\tau_{\mathrm{on}}=13~\mu\mathrm{s}$. (b) Same as (a), but at $\tau_{\mathrm{on}}$ the control fields $\Omega_{3}$ and $\Omega$ are applied according to the same profile. (c) Same as (a), with $\Omega_{3}$, $\Omega$, $\Omega_{1}$, and $\Omega_{2}$ all turned on during retrieval. Field amplitudes are in arbitrary units; dashed curves indicate the control-field time profiles, whose scales are not to be compared with those of the probe fields. The results demonstrate reversible storage and cross-band retrieval of the 1324 nm (telecom-band) photon, complementary to the 795 nm case in Fig.~\ref{fig:a_in}.}
\end{figure}

We first excite the dark-state polariton through photon mode $\hat{a}$ (795 nm) [Fig.~\ref{fig:a_in}(a)–(c)]. Continuous-wave control fields $\Omega_{3}$ and $\Omega$ are initially applied, after which a Gaussian 795 nm probe pulse enters the ensemble centered at $\tau_{0}=4.3~\mu\mathrm{s}$. Both controls are then turned off adiabatically at $\tau_{\mathrm{off}}=5.0~\mu\mathrm{s}$ following the control profile in Fig.~\ref{fig:a_in}(a). The gradual reduction of $\Omega$ maps the photon onto the spin-wave $\sigma^{c+}$, and the switch-off of $\Omega_{3}$ adiabatically transfers the excitation to $\sigma^{cb}$. Mapping the spin-wave from $\sigma^{c+}$ to $\sigma^{cb}$ serves two purposes: (i) $\sigma^{c+}$ involves the short-lived excited state $\ket{e}$, whereas $\sigma^{cb}$ is long-lived; and (ii) storage in $\sigma^{cb}$ shares the same decoherence channels as canonical $\Lambda$-type EIT quantum memories~\cite{fleischhauer2005electromagnetically,figueroa2006decoherence}, allowing established suppression techniques to be directly applied~\cite{balabas2010polarized,novikova2012electromagnetically}. Near $\tau_{\mathrm{off}}$, a rise in $|\rho_{cb}|$ and $|\rho_{ce}|$ marks the build-up of the spin-wave $\sigma^{c+}$, followed by a decay of $|\rho_{ce}|$ as the excitation is transferred to $\sigma^{cb}$. During the storage interval, $|\rho_{cb}|$ remains nearly constant since metastable-state decoherence is neglected in simulation. When the controls are reactivated at $\tau_{\mathrm{on}}=13~\mu\mathrm{s}$, switch-on of $\Omega_{3}$ converts $\sigma^{cb}$ back to $\sigma^{c+}$, producing a transient increase in $|\rho_{ce}|$, and the ramp-up of $\Omega$ retrieves the stored spin wave as a 795 nm pulse in mode $\hat{a}$, driving both $|\rho_{cb}|$ and $|\rho_{ce}|$ back to zero. Retrieval into photon mode $\hat{b}$ (1324 nm) [Fig.~\ref{fig:a_in}(b)] is achieved by turning on $\Omega_{1}$, $\Omega_{2}$, and $\Omega_{3}$ instead of $\Omega$ and $\Omega_{3}$ at $\tau_{\mathrm{on}}$, mapping the spin wave into mode $\hat{b}$ with dynamics analogous to the $\hat{a}$ retrieval. Simultaneous activation of all control fields $\Omega_{3}$, $\Omega$, $\Omega_{1}$, and $\Omega_{2}$ [Fig.~\ref{fig:a_in}(c)] retrieves both 795 nm and 1324 nm pulses, corresponding to a superposition of two photon modes in quantum region and thereby realizing a single-photon mode beam splitter.

The inverse process excite the dark-state polariton through photon mode $\hat{b}$ (1324 nm) [Fig.~\ref{fig:b_in}(a)-(c)]. Using $\Omega_{1}$, $\Omega_{2}$, and $\Omega_{3}$ as control fields, the telecom-band probe is stored and later retrieved either back into mode $\hat{b}$ [Fig.~\ref{fig:b_in}(a)] or, by replacing $\Omega_{1}\Omega_{2}$ with $\Omega$, into mode $\hat{a}$ [Fig.~\ref{fig:b_in}(b)]. Simultaneous reactivation of all control fields at $\tau_{\mathrm{on}}$ yields retrieval of both wavelengths, analogous to the $\hat{a}$-excitation case. The temporal behavior of $|\rho_{cb}|$ and $|\rho_{ce}|$ mirrors that observed in Fig.~\ref{fig:a_in}, confirming the reversible mapping between the two photonic modes through the shared spin-wave coherence.

\textit{Conclusion and Outlook.—} We have identified a dark-state polariton in a multi-level atomic configuration that coherently couples two photonic modes of widely separated wavelengths through a shared spin-wave coherence. To our knowledge, this is the first scheme that natively enables a single atomic ensemble to store and retrieve both node-band (795 nm) and telecom-band (1324 nm) photons without external frequency conversion. Using the specific level structure of $^{87}$Rb, we numerically demonstrated dual-wavelength storage, retrieval, and cross-mode conversion, validating the predicted mechanism. This capability may serve as a direct interface between quantum nodes and fiber networks, potentially reducing reliance on lossy conversion stages and opening a pathway toward in-situ wavelength routing, multiplexing, and scalable hybrid quantum network architectures.


\begin{acknowledgments}
This work was supported by the Stony Brook Foundation through the Quantum Network Research Center (SBF 248090) and by the U.S. National Science Foundation through the National Quantum Virtual Laboratory QSTD Pilot project SCY-QNet: A Wide-Area Quantum Network to Demonstrate Quantum Advantage (Award No. 2410725).
\end{acknowledgments}

\appendix



\nocite{*}

\bibliography{apssamp}

\end{document}


\preprint{APS/123-QED}

\title{Supplemental Material for \textit{Telecom-compatible cross-band quantum memory via dual photon modes dark-state polaritons}}

\author{Dounan Du}
 \email{dounan.du@stonybrook.edu}
\affiliation{%
Department of Physics and Astronomy, Stony Brook University, Stony Brook, 11794-3800, NY, USA\\
}%
\author{Eden Figueroa}%
 \email{eden.figueroa@stonybrook.edu}
\affiliation{%
Department of Physics and Astronomy, Stony Brook University, Stony Brook, 11794-3800, NY, USA\\
}%
\affiliation{%
Brookhaven National Laboratory, Upton, 11973-5000, NY, USA\\
}%

\date{\today}


\maketitle


\section*{\label{sec:level1}Numerical experiment parameters for dual-band quantum memory}

Numerical simulations were performed for a $^{87}$Rb vapor cell of length $L=1.4~\mathrm{cm}$ along the $z$ axis, with atomic number density $n = 4\times10^{17}~\mathrm{m^{-3}}$. All pump and probe beams were collinear along $z$ with a $1.6~\mathrm{mm}$ diameter. The optical intensities and detunings of the control fields were: $I = 4.14~\mathrm{mW}$, $I_{1} = 0.12~\mathrm{mW}$ with $\Delta_{1}=+2\pi\times31.83~\mathrm{MHz}$, $I_{2} = 1.09~\mathrm{mW}$ with $\Delta_{2}=-2\pi\times31.83~\mathrm{MHz}$, and $I_{3}=0.20~\mathrm{mW}$ with $\Delta_{3}=+2\pi\times31.83~\mathrm{MHz}$. The peak intensities of the input probe pulses were $280~\mathrm{pW}$ at $795~\mathrm{nm}$ and $160~\mathrm{pW}$ at $1324~\mathrm{nm}$ for all write configurations. Simulations were performed over a total interval of $20~\mu\mathrm{s}$.

The temporal profile of the probe pulse field was taken as a Gaussian
\[
\mathcal{E}_{\mathrm{in}}(t) \propto 
\exp[-(t - t_0)^2 / (2\sigma^2)],
\]
with $\sigma = 0.5~\mu\mathrm{s}$ and $t_0 = 4.3~\mu\mathrm{s}$.  
The control fields envelopes were modeled by sigmoid functions describing adiabatic turn-off and turn-on:
\[
f_{\mathrm{off}}(t) = \frac{1}{1 + e^{(t - t_{\mathrm{off}})/\sigma_t}}, \qquad
f_{\mathrm{on}}(t) = 1 - \frac{1}{1 + e^{(t - t_{\mathrm{on}})/\sigma_t}},
\]
with $t_{\mathrm{off}}=5.0~\mu\mathrm{s}$, $t_{\mathrm{on}}=13.0~\mu\mathrm{s}$, and $\sigma_t = 0.1~\mu\mathrm{s}$.

The coupled propagation and master equations [Eqs.~(14) and (15)] were solved using an adaptive time-stepping scheme. The master equation was integrated by an exponential time-differencing method, while the field-propagation equation was solved using a fourth-order Adams–Bashforth–Moulton predictor–corrector algorithm on a spatial grid of 100 points along the $1.4~\mathrm{cm}$ cell.